\documentclass[9pt,twocolumn]{article}

\usepackage{graphicx,epstopdf,caption}
\usepackage{amssymb}
\usepackage{siunitx}
\usepackage{breakcites}
\usepackage{authblk}
\usepackage[a4paper, total={7in, 9in}]{geometry}
\usepackage{floatrow}
\usepackage{balance}
\usepackage{caption}

\newcommand{\be}{\begin{equation}}
\newcommand{\ee}{\end{equation}}
\newcommand{\bee}{\begin{eqnarray}}
\newcommand{\eee}{\end{eqnarray}}
\newcommand{\I}{ I }
\newcommand{\J}{ J }
\newcommand{\C}{ C }
\newcommand{\W}{ W }
\newcommand{\F}{ \mathcal{F} }
\newcommand{\A}{ \mathcal{A} }

\newcommand{\comment}[1]{{#1} }

\title{Viscosity of Cohesive Granular Flows}
\author{Matthew Macaulay and Pierre Rognon}
\affil{Particles and Grains Laboratory, School of Civil Engineering, The University of Sydney, NSW 2006 Sydney Australia}
\date{}

\begin{document}

\twocolumn[
\begin{@twocolumnfalse}
	\maketitle
		\begin{abstract}
		Cohesive granular materials such as wet sand, snow, and powders can flow like a viscous liquid. However, the elementary mechanisms of momentum transport in such athermal particulate fluids are elusive. As a result, existing models for cohesive granular viscosity remain phenomenological and debated. Here we use discrete element simulations of plane shear flows to measure the viscosity of cohesive granular materials, while tuning the intensity of inter-particle adhesion. We establish that two adhesion-related, dimensionless numbers control their viscosity. These numbers compare the force and energy required to break a bond to the characteristic stress and kinetic energy in the flow. This progresses the commonly accepted view that only one dimensionless number could control the effect of adhesion. The resulting scaling law captures strong, non-Newtonian variations in viscosity, unifying several existing viscosity models. We then directly link these variations in viscosity to adhesion-induced modifications in the flow micro-structure and contact network. This analysis reveals the existence of two modes of momentum transport, involving either grain micro-acceleration or balanced contact forces, and shows that adhesion only affects the later. This advances our understanding of rheological models for granular materials and other soft materials such as emulsions and suspensions, which may also involve inter-particle adhesive forces.
		\end{abstract}
\end{@twocolumnfalse}
]

Continuum fluid mechanics models are of considerable interest to predict the dynamics of natural and industrial granular flows. 
However, they hinge on the knowledge of the shear viscosity of granular fluids. 

A robust scaling law for granular viscosity has been established for dry grains which have no contact adhesive force. By analogy with Newtonian fluids, the apparent granular viscosity $\eta$ was defined as the ratio of the shear stress $\tau$ and shear rate $\dot \gamma$: $\eta = \tau/ | \dot \gamma |$. A major breakthrough was the identification of a frictional constitutive law relating the bulk shear stress to the pressure $P$, $\tau = \mu P$. Like Coulomb friction, it involves a coefficient of friction $\mu$. The complexity of granular flows is rationalised by a unique friction law $\mu(\I)$ relating the friction coefficient to a single dimensionless number, called the inertial number $\I=d\dot \gamma \sqrt{\rho_g/P}$, involving the grain size $d$ and density $\rho_g$. This led to establishing a general scaling law for granular viscosity \cite{midi2004dense,da2005rheophysics,jop2006constitutive,forterre2008flows}:

\be \label{eq:visco}
\eta = \mu(\I) \frac{P}{|\dot \gamma|}.
\ee

\noindent This law captures complex non-Newtonian features of granular flows, including shear-thinning and a viscosity divergence when flows stop. Its domain of validity has been extended to granular materials submerged in a fluid of viscosity $\eta_f$, by introducing a second dimensionless number $J = \eta_f/P$ contributing to a friction law that became $\mu(\I,\J)$ \cite{rognon2011flowing,degiuli2015unified}. 

However, there is no consensus on extending this viscosity scaling to the vast range of granular materials featuring inter-granular adhesion. These typically include materials with grain size smaller than \SI{100}{\micro\meter} which tend to adhere via van der Waals surface interactions or wet grains of all sizes which stick via capillary forces. The difficulty in establishing a cohesive granular viscosity is that these modes of adhesion have different physical characteristics. Nonetheless, they all involve two elementary contact parameters: an adhesion force $f_0$ and an adhesion energy $w_0$. These correspond to the minimum force and energy needed to unstick two grains.

Adhesive forces are known to strongly affect the micro-structure of granular flows. They induce the formation of large clusters in the flow \cite{rognon2006rheophysics, rognon2008dense, macaulay2019shear}, which enhances the process of shear-induced dilation \cite{iordanoff2005numerical,khamseh2015flow,roy2017general,koeze2018sticky, shi2019steady,irani2014impact,khamseh2015flow}. These micro-structural changes generally coincide with an increase in friction. The usual approach to rationalise this effect is to express a friction law, which depends on the inertial number and a dimensionless number $C = \frac{f_0}{Pd^2}$. This compares the contact tensile strength $f_0$ to the typical force scale related to the pressure $P$. Various phenomenological expressions for $\mu(\I,\C)$ were introduced, which capture measurements made in many flow configurations and with different modes of adhesion \cite{rognon2006rheophysics, rognon2008dense,aarons2006shear, gu2014rheology, brewster2005plug, metayer2010electrically, berger2016scaling, roy2017general,vo2020additive}. The consensus is that the friction is not significantly enhanced at low values of $C \lesssim 1$ and then increases with $C$. This transition delineates between a non-cohesive and a cohesive flow regime. However, several observations indicate that the viscosity of cohesive granular flows is not solely controlled by the number $C$: at a fixed value of $C$, grains that are softer or more dissipative lead to larger friction \cite{mandal2020insights}, while a faster shear rate induces a decrease in friction \cite{aarons2006shear, gu2014rheology}.

We propose that the apparent discrepancies in existing scaling laws for cohesive granular viscosity reflect the existence of different cohesive flow regimes controlled either by the strength $f_0$ or the energy $w_0$ of adhesive forces. In this paper, we assess this assumption using a set of simulations of steady plane shear flows of adhesive grains, with varying combinations of these two adhesive parameters. The goal of these numerical experiments is to simultaneously measure the material viscosity and to identify the internal mechanisms of momentum transport controlling it. 

\begin{figure}[!tb]
	\centering
	\includegraphics[width=\linewidth]{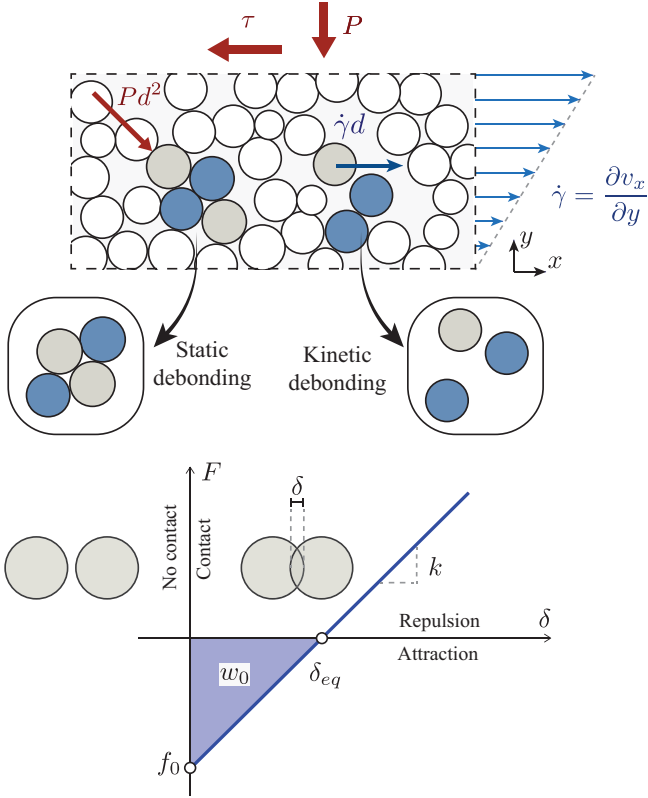}
	\caption{{\bf Simulated shear flows of cohesive grains.} (a) Plane shear flow under constant normal stress $P$ and strain rate $\dot \gamma$. The dashed lines represents periodic boundaries and the blue arrows illustrate the linear velocity profile that usually develops; the simulated system is bi-dimensional and its size is approximately $100d\times100d$. The insets illustrate two mechanisms of de-bonding whereby a pair of contacting grains (blue) is pulled apart by either a static force or a kinetic impact. 
	(b) Adhesive contact model: normal, quasi-static force $F$ between two grains including a linear elastic repulsion and a constant adhesion $f_0$.}
	\label{fig1}
\end{figure}

\section*{Methods}

\subsection*{Measuring viscosity in simulated plane shear flows}

To measure the viscosity of cohesive granular flows, we used a discrete element method to simulate bi-dimensional plane shear flows of $10\,000$ sticky grains. This section presents the key physical properties of the grains and of the flows configuration. Details of the simulated system are given in the electronic supplementary information.

The plane shear flow configuration is illustrated in figure~\ref{fig1}a. It involves prescribing both the normal stress $P$ and the shear rate $\dot \gamma$. This is achieved using Lees-Edwards periodic boundary conditions to avoid any solid boundaries and subsequent flow heterogeneities they would induce~\cite{rognon2015long}. The normal stress is controlled by allowing the cell height to expand or contract during the flow. The advantage of this configuration is to produce steady flows characterised by a homogeneous shear rate $\dot \gamma$, normal stress $P$ and shear stresses $\tau$. It thus enables us to prescribe a single value of $\I$ and to directly measure the resulting friction coefficient $\mu(\I)$. 

Grains are disks of diameter $d\pm20\%$. This slight polydispersity is introduced to prevent shear-induced crystallisation. Grains interact via direct contacts with their neighbours. Contact forces are comprised of friction, elastic repulsion, dissipation and adhesion. The inter-granular friction coefficient is $0.5$ and the normal, non-cohesive coefficient of restitution is $0.5$ in all simulations. The effect of these parameters on the flow properties can be found in \cite{da2005rheophysics} for non-cohesive flows and in \cite{singh2015role,mandal2020insights} for cohesive flows.

The elastic repulsion follows a Hookean law $F^{e}(\delta) = k \delta$ where $\delta$ measures the interpenetration of two contacting grains and $k$ is an elastic stiffness parameter. An elementary dimensionless number measuring the softness of the grains is:

\be
K = \frac{Pd}{k}.
\ee

The adhesion model is chosen to be the simplest: a constant attractive normal force $F^{a}(\delta) = -f_0$ is active while two grains are in contact. The resulting normal force between two immobile contacting grains $F=F^{e}+F^{a}$ is illustrated in figure \ref{fig1}b. It is characterised by an equilibrium position $\delta_{eq} = \frac{f_0}{k}$, a maximum strength $f_0$ and an adhesion energy $w_0 =\frac{1}{2}k \delta_{eq}^2 = \frac{f_0^2}{2k}$. Accordingly, two elementary dimensionless numbers characterise the intensity of adhesion in such flows:

\be
C = \frac{f_0}{Pd^2} \quad \text{and} \quad W = \sqrt{\frac{w_0}{\frac{1}{2}m \dot \gamma^2 d^2}} = \frac{CK^{\frac{1}{2}}}{I}. 
\ee

\noindent The cohesion energy number $W$ compares the cohesion energy to the characteristic kinetic energy of grains colliding at a relative velocity $\dot \gamma d$. This dimensional analysis points out two processes by which a pair of contacting grains may be pulled apart, which we called \textit{static} and \textit{kinetic} de-bonding. An illustration of these processes is shown in figure \ref{fig1}a. Static de-bonding is likely at low cohesion strength ($C \lesssim1$), when sustained forces of magnitude $Pd^2$ exceed the adhesion strength $f_0$. Kinetic de-bonding is likely at low values of cohesion energy ($W \lesssim1$), when colliding grains have enough kinetic energy to overcome the adhesion energy $w_0$.

%One could therefore expect that adhesion might not significantly affect the flow properties if either $C$ or $W$ have a low values. Reciprocally, the condition for adhesion to affect the flow properties would be to have a large enough value of both $C$ and $W$.

\begin{figure}[!ht]
	\begin{center}
		\includegraphics[width=0.865\linewidth]{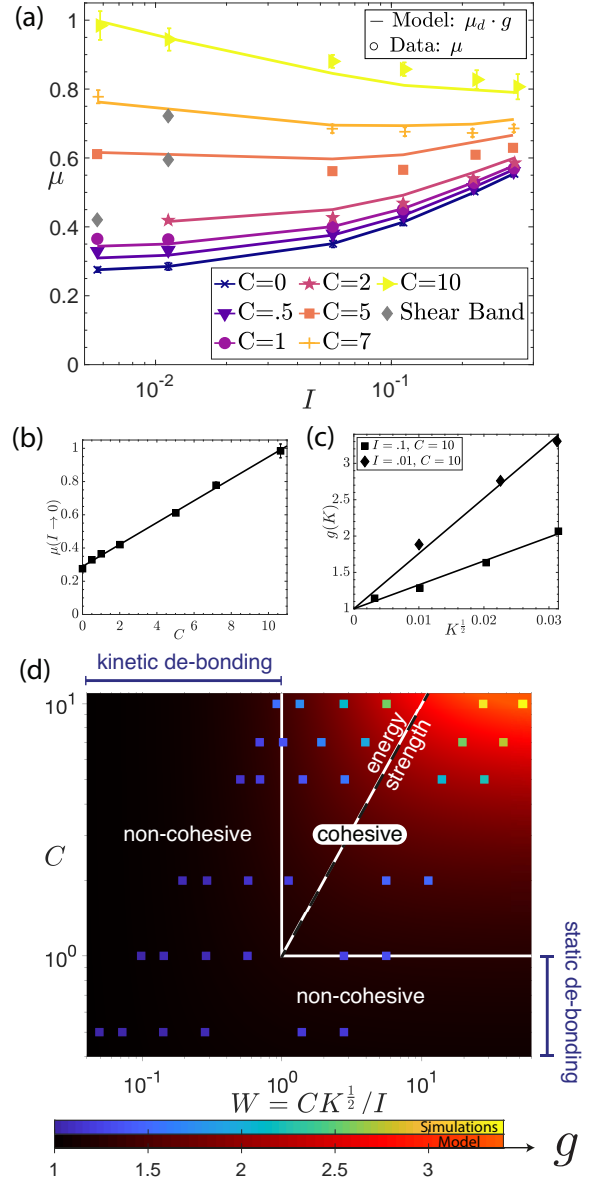}
		\caption{{\bf Viscosity measurements.}
			 (a) Friction law $\mu(\I)$ measured at different cohesion strengths $\C$ (the stiffness number is fixed at $K=10^{-3}$ in a,b,d): markers are the simulation results and solid lines represent the proposed phenomenological model Eq.~(\ref{eq:musplit},\ref{eq:g}); some flows developed heterogenous shear state in the form of a persistent shear bands (see the ESI). 
			 (b) Quasi-static friction $\mu(I\to0)$: markers represent the friction measured at the lowest value of the inertial number $\I=5\times10^{-3}$; the solid line is the best linear fit of this data: $\mu(\I\to 0) = 0.29 + .066 C$.
			(c) Friction enhancement factor $g$ scaling with particle softness; the lines represents the proposed model in Eq.~(\ref{eq:musplit},\ref{eq:g}).
			(d) Viscosity regimes: markers colours indicate measured values of the friction enhancement factor $g(\I,\C)= \mu/\mu_d$ using Eq.~(\ref{eq:mud}); background colour is the model in Eq.~(\ref{eq:g}) calculated for various value of $W$ corresponding to different combinations of $\C$ and $\I$; the white lines at $C=1$ and $W=1$ delineates the cohesive and non-cohesive regimes; the dashed line $C=W$ delineates the cohesive-strength and cohesive-energy regimes. 
			\label{fig2}}
	\end{center}
\end{figure}

\section*{Results}
\subsection*{Cohesion strength and energy control viscosity}

In order to quantify the effect of adhesive forces of granular viscosity, we simulated a series of plane shear flows, selecting a value of inertial number, cohesion-strength number and stiffness number in the ranges $5.10^{-3} \leqslant \I \leqslant 0.3 $, $0 \leqslant \C \leqslant 10$, and $10^{-5} \leqslant K \leqslant 10^{-3}$. \comment{With no cohesion, these values of $K$ are small enough to not affect the rheology \cite{da2005rheophysics}; the rheology of softer grains is discussed in \cite{singh2015role,roy2017general,shi2019steady}.} The choice of these three numbers determines the value of the cohesive-energy number $W(C,I,K)$, which ranges from $0$ to $63$. 

Once steady and homogeneous shear flow developed, the friction $\mu$ was measured by averaging the stresses across the entire shear cell during $15$ shear deformation. Figure \ref{fig2}a shows the resulting set of friction laws $\mu(\I,C)$, obtained at a fixed stiffness $K=10^{-3}$. At low cohesion strength ($C \lesssim 1$), the friction law is similar to that of non-cohesive grains ($C=0$). It follows the established empirical law \cite{da2005rheophysics,midi2004dense,jop2006constitutive,forterre2008flows}:

\be \label{eq:mud}
\mu_{d}(\I) = \mu_0 + \frac{\mu_2-\mu_0}{1+\I_0/\I},
\ee

\noindent with $\mu_0=0.266\pm0.001$, $\mu_2=0.830\pm0.014$, and $\I_0=0.0316\pm0.015$ (standard errors given throughout text and figures). This friction law captures a \textit{shear-rate strengthening} behaviour whereby higher inertial numbers lead to higher friction $\mu$.
 
In contrast, the friction may be significantly increased at higher cohesion strengths ($C \gtrsim 1$). Then, the friction law $\mu(\I)$ measured at a constant value of $C$ are qualitatively different from their non-cohesive counterpart $\mu_{d}(\I)$. Strikingly, a phenomenon of \textit{shear-rate weakening} occurs at the highest values of $C$, by which the friction coefficient decreases as the inertial number is increased. 

We propose a phenomenological scaling law to capture the measured cohesive friction $\mu(\I,C)$ that involves not only $C$ but also the cohesion energy $W$. We choose the following functional form: 

\bee
\mu &=& \mu_{d}(\I) g(W,\I) \label{eq:musplit}\\
g(W,I) &=& 1 + b \frac{W}{1+\I_1/\I}, \label{eq:g}
\eee

\noindent involving two numerical constants $b$ and $\I_1$. Figure \ref{fig2}a shows that these expressions closely capture all the friction laws $\mu(\I)$ measured at various levels of $C$ using $b=0.527\pm0.021$ and $\I_1 = 0.062\pm0.003$. In particular, they capture the transition from shear-rate strengthening to shear-rate weakening behaviour. 

This chosen functional form is not derived from a physical process and other choices could possibly capture the data equally well. Nonetheless, it conveniently highlights three ways cohesion might affect granular viscosity. Firstly, any effect of adhesion is included into the term $g(W,\I)$, which may be seen as a friction enhancing factor induced by adhesion. This readily distinguishes a regime of low cohesion ($g(W,\I)\approx 1$) where friction is unaffected and similar to that of non-cohesive grains, to a regime of high cohesion ($g(W,\I)> 1$) where adhesion significantly enhances friction. This criterion delineates the \textit{non-cohesive} and \textit{cohesive} flow regimes.

\begin{figure*}[!ht]
	\centerline{\includegraphics[width=\textwidth]{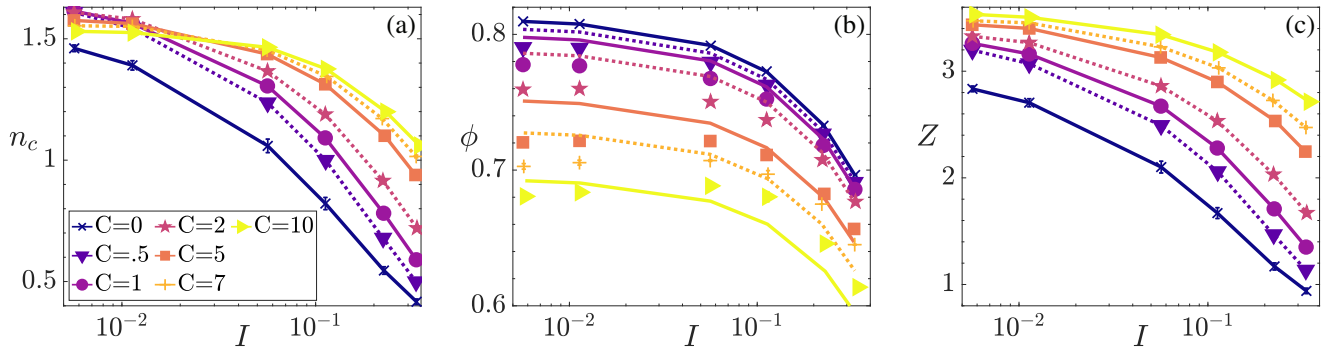}}
	\caption{{\bf Microstructure of cohesive flows.} (a) Contact density $n_c$; (b) solid fraction $\phi$; and (c) coordination number $Z$ at a function of the flows' inertial number $I$ and cohesion strength $C$. Grain stiffness is constant $K=10^{-3}$. Markers are numerical results; lines in (b) are the best linear fits obtained using Eq.~(\ref{eq:phi}) and Eq.~(\ref{eq:phid}).}
	\label{fig3}
\end{figure*}

The law in Eq.~(\ref{eq:g}) quantifies the friction enhancement in the cohesive flow regime and indicates a transition between two cohesive regimes. At low inertial number, it reduces to

 \be \label{eq:gC}
 g (\I \ll I_1)\approx 1+ b C \frac{ K^{\frac{1}{2}} }{\I_1}.
 \ee 

\noindent The effect of adhesion on friction is then independent from the cohesion energy $W$ and from the inertial number. At a fixed grain stiffness, it is solely controlled by the cohesion strength $C$. Figure \ref{fig2}b confirms that in the limit of $\I \to 0$, the friction $\mu$ increases linearly with $C$ for a fixed value of stiffness $K$. We call this flow regime \textit{cohesive-strength}. This linear increase of the friction with $C$ is consistent with several existing results \cite{rognon2006rheophysics, rognon2008dense,iordanoff2005numerical,khamseh2015flow,roy2017general,vo2020additive}. Furthermore, this friction scaling with the stiffness number $K$ is qualitatively consistent with the observation that softer grains (larger $K$) lead to an increase in friction \cite{mandal2020insights}.

In contrast, at high inertial numbers Eq.~(\ref{eq:g}) reduces to: 
\be \label{eq:gW}
g(\I \gg I_1) \approx 1+ b W = 1+ b C \frac{K^{\frac{1}{2}}}{\I}.
\ee

\noindent The effect of adhesion on friction is then controlled by the adhesion energy $W$. We call this flow regime \textit{cohesive-energy}. The scaling in Eq.~(\ref{eq:gW}) is qualitatively consistent with the observations that softer grains lead to an increase in friction \cite{mandal2020insights}\comment{; that at high inertial numbers, the scale of kinetic energy is important \cite{gonzalez2014free}, and} may induce a decrease in friction for cohesive flows with a fixed value of $C$ \cite{aarons2006shear, gu2014rheology}, as with granular suspensions \cite{perrin2019interparticle}. Figure \ref{fig2}b shows the friction $\mu$ measured in flows performed at a fixed value of $\I$ and $C$, and with differing stiffness number $K$. It confirms that $g$ linearly increases with $K^{\frac{1}{2}}$.

According to Eq.~(\ref{eq:g}), the transition between the cohesion-strength and cohesion-energy regimes occurs at $\I\approx \I_1$. The cohesion energy is then $W = C K^{\frac{1}{2}}/I_1$. Our results indicate that the ratio $K^{ \frac{1}{2}} /\I_1$ is of the order of unity ($I_1=0.063$, and $K^\frac{1}{2}= 0.03$). As a first order approximation, we then propose a regime transition for $C=W$. The cohesive-strength regime develops for $C<W$, and the mechanism of static de-bonding is typically the weakest way to pull grains apart and therefore controls the effect of adhesion. Reciprocally, the cohesive-energy regime develops for $C>W$.

Figure \ref{fig2}d maps the occurrence of these three regimes (non-cohesive, cohesive-strength and cohesive energy), by plotting the measured friction enhancement $g = \frac{\mu}{\mu_d}$ as a function of the cohesion strength $C$ and energy $W$, keeping the stiffness $K=10^{-3}$ constant. It confirms that $g$ becomes significantly larger than one only if both cohesion strength and energy are greater than one. This progresses the view that cohesive and non-cohesive regimes may be delineated by only one dimensionless number. 

\comment{Section 3 of the ESI presents a detailed analysis of previously proposed model of friction laws. It provides evidence that previous models successfully captured either cohesive-strength or the cohesive-energy regime, but not both. The rheology proposed here distinguishes these two cohesive regimes, enabling one to reconcile the apparent differences in these models.}

\subsection*{Internal mechanisms of momentum transport}

We now seek to understand the behaviour of cohesive viscosity by establishing the elementary mechanisms of momentum transport controlling it. As a starting point, we consider the basic definition of stress in granular materials in terms of individual contact force $f_c$ and contact branch vector $l_c$, which joins the centre of a grain to the point of contact. The average stress tensor $\sigma$ is given by: $\sigma^{\alpha \beta} = n_c \langle \textnormal{sym}( f^{\alpha}_c l^{\beta}_c) \rangle$. $n_c$ is the contact density or number of contacts per unit volume; $\textnormal{sym}(f^{\alpha} l^{\beta})$ is the symmetric part of an individual contact moment tensor $f^{\alpha}_c l^{\beta}_c$. $\langle \cdot \rangle$ is the averaging operator including all contacts in the flow. As we observed that the moment tensor is symmetrical, the shear stress may be expressed as $\tau = n_c \langle f^y l^{x}\rangle = n_c \langle f^x l^{y}\rangle$. 

The shear stress represents the flux of momentum through the flow. This expression relates it to individual contacts, which readily bridges the continuum scale to micro-dynamical properties. However, it does not reveal any specific mechanism of momentum transport. To highlight them, we use a mathematical expansion of this expression introduced in \cite{macaulay2020two}: $\langle f^y l^{x}\rangle = \langle f^y \rangle \langle l^{x}\rangle + \hat{ f}^y \hat{l}^x \mathcal{C}$, where $\hat{f}^y$ and $\hat{l}^x$ are the standard deviation of the force and branch vector components, and $\mathcal{C}$ is their correlation. Since the average force $\langle f^y \rangle$ is null in steady flows, this reduces to $\langle f^y l^{x} \rangle = \hat{ f}^y \hat{ l}^x \mathcal{C}$. We further observed that the term $\hat{ l}^x$ is proportional to the grain size: $\hat{ l}^x \approx \alpha_l d$ with a constant $\alpha_l\approx0.71$ that we found to be virtually independent on adhesion and inertial number. This leads to the following expression: 

\be \label{eqn:mu_df}
\mu = \frac{\tau}{P} = \beta \F; \quad \F = \frac{\hat{ f}^y}{Pd^2}; \quad \beta = n_c d^3 \alpha_l \mathcal{C}.
\ee

\noindent Hereafter, we refer to the dimensionless number $\F$ as \textit{force fluctuations}. This expression shows how macroscopic friction, and hence granular viscosity, is directly related to three measurable microscopic quantities: the contact density, the force fluctuations and the force correlation. The following presents how adhesion affects each of them.

\begin{figure*}[!ht]
	\begin{center}
		\centerline{\includegraphics[width=\linewidth]{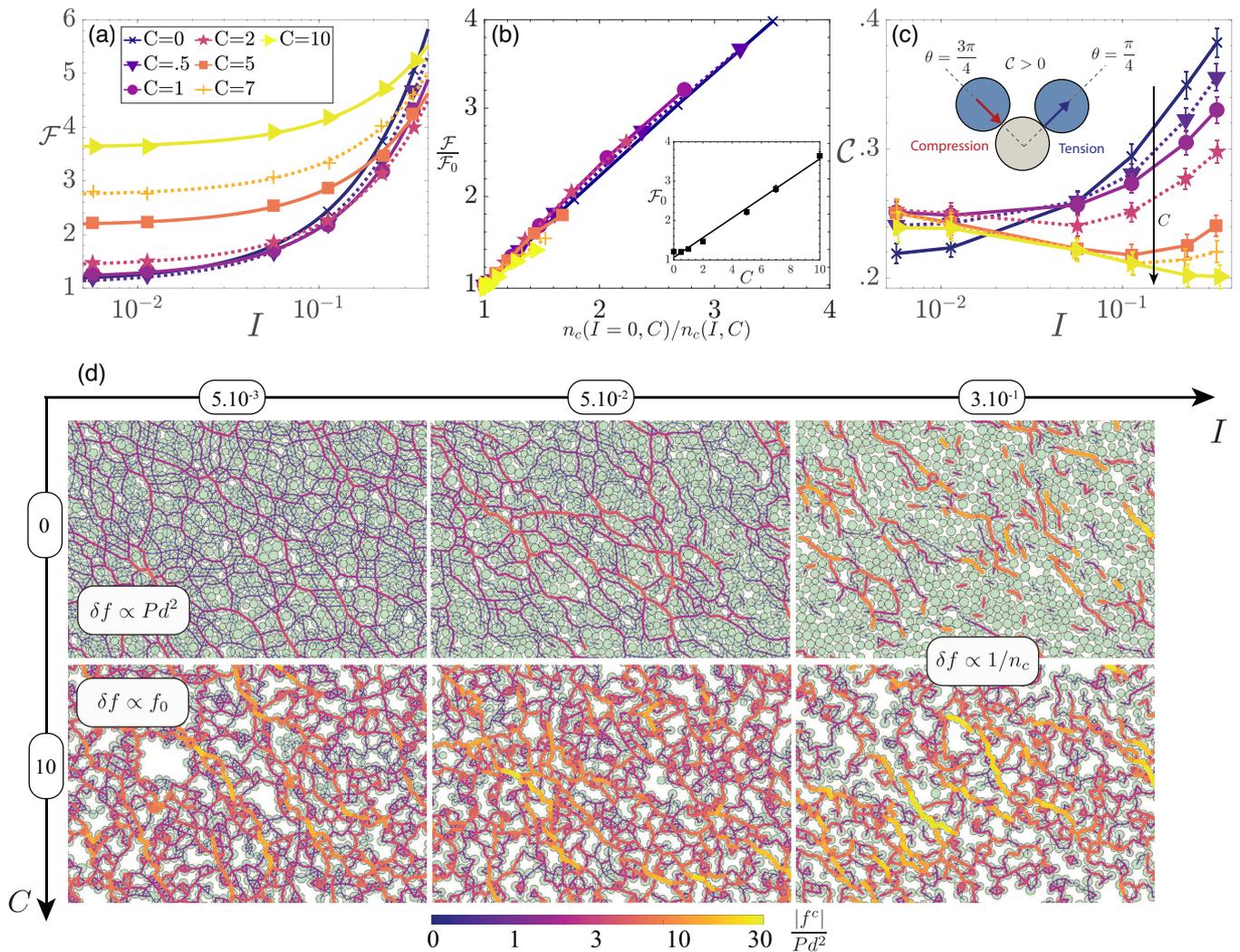}}
		\caption{{\bf Contact forces in cohesive flows at different inertial number and cohesion strength} (constant grain stiffness $K=10^{-3}$). (a) Force fluctuations $\F$ measuring the standard deviation of contact force y-component. (b) Force fluctuations as a function of the contact density; $\F_0 = \F(\I\to 0)$ and $n_c(\I \to 0)$ are the quasi-static limit of the force fluctuations and contact density, which are approximated here by the value measured at the lowest inertial number $\I=5\times 10^{-3}$. The inset shows how this quasi-static limit of force fluctuations increases linearly with cohesion; the line represents the best linear fit: $\F_0=1+0.26C$. (c) Correlation between contact force and branch vector cross-components. The inset illustrate two contact orientations that maximise this term. (d) Snapshots of contact forces $f_c$ taken during steady and homogeneous shear flows. Lines join pairs of grains in contact with a width and colour representing the magnitude of the contact force on a logarithmic scale. The ESI contains corresponding movies.}
		\label{fig4}
	\end{center}
\end{figure*} 

\begin{figure*}[!ht]
	\begin{center}
		\centerline{\includegraphics[width=\linewidth]{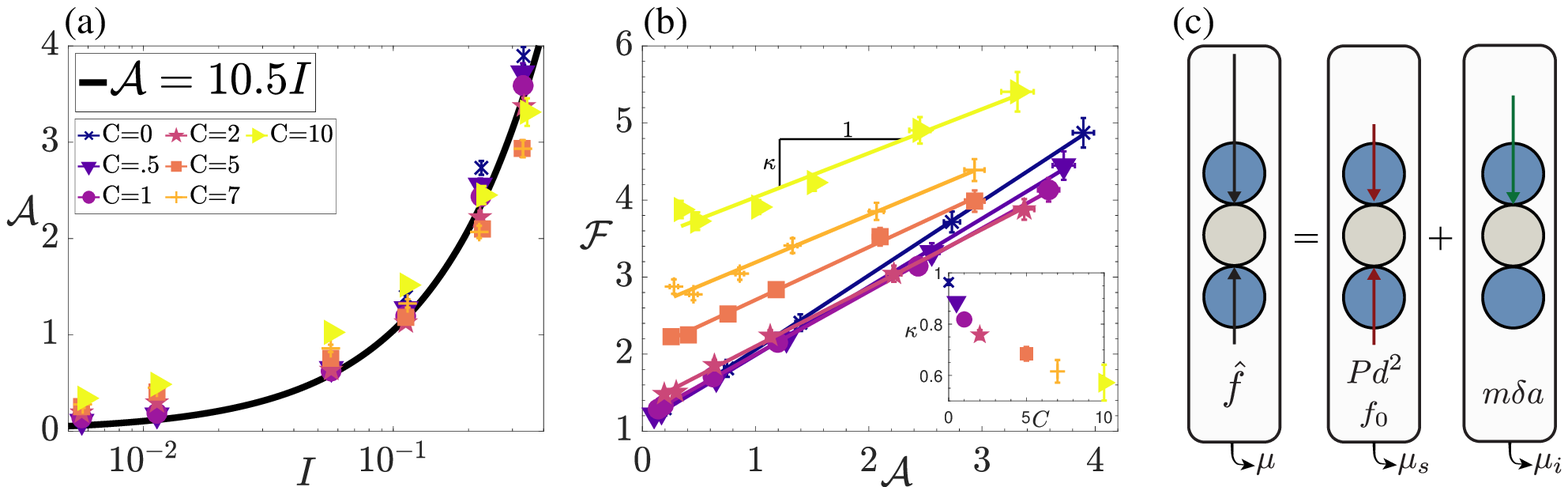}}
		\caption{{\bf Grain micro-acceleration} (constant grain stiffness $K=10^{-3}$). (a) Acceleration fluctuations at different inertial number and cohesion strength; the black line shows the fitted function $\A = 10.5 \I$. Relation between fluctuation in grain acceleration $\A$ and contact force $\F$; markers are the simulated data and lines are the best linear fit using (\ref{eqn:df_A}), obtained for a value of $ \F_0(C)$ given by Eq.~(\ref{eq:dF0}) and a value of $\kappa$ shown on the inset. (c) Illustration of the decomposition of the contact force fluctuations into a balanced and an unbalanced part, controlling the static and inertial components of the friction $\mu$ respectively.}
		\label{fig5}
	\end{center}
\end{figure*} 

\subsubsection*{Adhesion enhances contact density} 
This counter-intuitive observation is evidenced in figure~\ref{fig3}a: at a given inertial number, the contact density of a cohesive flow is equal or larger than in a non-cohesive flow. This is unexpected as cohesion simultaneously enhances the process of shear-induced dilation. This discord reflects the potential for granular matter to form fragile force networks \cite{bi2011jamming} and the fact that the solid fraction $\phi$ decreases when the inertial number is increased. Figure~\ref{fig3}b shows that the solid fraction is significantly lower in cohesive flows than in non-cohesive flows, as previously observed \cite{shi2019steady, koeze2018sticky, roy2017general, rognon2008dense, rognon2006rheophysics}. In a first approximation, we propose to capture these variations by the following dilatancy law: 

 \bee
 \phi &=& \phi_{d}(\I) h(C );\label{eq:phi}\\
 \phi_d(\I) &\approx& \phi_0 - b_\phi \I; \;\; h(\C) \approx 1 - c_\phi C \label{eq:phid}.
 \eee

\noindent $\phi_{d}(\I)$ is the solid fraction of a non-cohesive flow and $h(\C)<1$ is a cohesion-driven reduction factor. Our data are captured by the following set of constants: $\phi_0=0.812\pm0.001$, $b_\phi=0.429\pm0.002$, $c_\phi = 0.014\pm0.001$. These two linear functions are phenomenological and other functional forms could be used to capture the data. 

Having a higher contact density with a lower solid fraction is made possible by an increase in coordination number, which measures the average number of contacts per grain. The relation between these three quantities is $n_c=\phi Z$. Figure~\ref{fig3}c shows that flows at high cohesion-strength, for instance $C=10$, can keep a high coordination number even at large inertial number. 
Nonetheless, even flows with the highest cohesion-strength exhibit a drop in coordination number at high inertial numbers, which coincides with the onset of the cohesive-kinetic regime. This decrease may therefore be attributed to the process of kinetic de-bonding, which becomes effective at separating grains in this regime.

\subsubsection*{Adhesion enhances force fluctuations}

Figure \ref{fig4}a shows that the force fluctuations are generally enhanced in the presence of adhesive forces. This enhancement, which is very pronounced at low inertial numbers, vanishes at high inertial numbers. This can be understood by estimating the range of force magnitude that contacts transmit. This is illustrated in the contact network snapshots shown in figure \ref{fig4}d.

With no cohesion, the high solid fraction and coordination number enable a contact network to percolate through the packing and connect virtually every grain. This means that most grains and contacts are involved in carrying the stresses, which scale with $P$. Accordingly, the force fluctuations is of the order of $\F \approx 1$. As the inertial number increases, more and more grains are disconnected from the network and have no contacts. The stresses are then supported by fewer bearing contacts, each of which must transmit a larger force. This qualitatively explains why the force fluctuations increases with $\I$.

With cohesion, the contact force magnitude is no longer driven by the normal stress, as it can vary from $0$ to the adhesive strength $f_0$ while pairs of grains are pulled apart. At low inertial numbers, this yields the following scaling for force fluctuation: $\mathcal{ F} \propto C$. At higher inertial numbers, as with non-cohesive grains, a process of stress concentration onto fewer bearing contacts leads to an increase in $\F(\I)$. 

We propose to capture the evolution of force fluctuations with the following model:

\bee
\F_0(C) &=&\F(\I \to 0) \approx 1+ \alpha_c C \label{eq:dF0}\\
\F(\I,C)&=& \F_0(C) \frac{n_c(\I=0,C)}{n_c(\I,C)}\label{eq:dF}.
\eee

\noindent Figure \ref{fig4}b confirms the inverse proportionality between the force fluctuations and the contact density $\F \propto n_c^{-1}$. It also confirms the linear increase of $\F_0(C)$ as in Eq.~(\ref{eq:dF0}), with fitted constant $\alpha_c = 0.26\pm0.01$. This force fluctuation $\F_0$ provides a coarse metric of correlated motions close to the jamming transition~\cite{corwin2005structural}.

\subsubsection*{Adhesion reduces force correlations}

Adhesion tends to enhance the contact density and the contact force fluctuations, which both enhance the friction. In contrast, figure \ref{fig4}c shows that it leads to a drop in the correlation $\C$ at high inertial numbers, which hinders the increase in friction. 
 
The force correlation term $\mathcal{C}$ measures how efficient contact forces are, on average, at supporting the shear stress~\cite{macaulay2020two,majmudar2005contact}. Considering contacts carrying a normal force only, the correlation term is driven by the contact orientation $\theta$, as $\mathcal{C} \propto \langle f_c \cos\theta \sin\theta \rangle / \hat{f}$, where $f_c$ is the value of the normal contact force which can be negative if the contact is in compression or positive if it is in tension. Ignoring any covariance between force magnitude and orientation angle gives an approximation for this expression $\mathcal{C} \propto \langle \cos\theta \sin\theta \rangle$, depending only on the orientation angle. Accordingly, two orientations that maximise the efficiency of such contacts in terms of contribution to the shear stress $\tau$ are: $\theta = 3\pi/4$ for compressive contacts and $\theta = \pi/4$ for contacts in tensions, as illustrated in figure \ref{fig4}c. Any other orientation would lessen $\mathcal{C}$, and an isotropic force distribution would lead to $\mathcal{C}=0$ and no shear stress. \comment{This result is consistent with the finding of an increase in force anisotropy by cohesion \cite{singh2014effect} }.

Without cohesion, all contacts are in compression. As the inertial number is increased, figure \ref{fig4}d shows that the remaining contacts are preferentially aligned at $\theta = 3\pi/4$, which drives the increase in correlation $\mathcal{C}$. With cohesion, these compressive contacts also develop at high inertial numbers. However, many contacts with seemingly random orientations remain, which contribute to lessening $\mathcal{C}$. This means that not all contacts actively contribute to the transport of momentum.

\subsubsection*{Adhesion enhances only one mode of momentum transport}

Expressing the friction $\mu$ in terms of force fluctuations in (\ref{eqn:mu_df}) enables us to further the analysis of elementary mechanisms of momentum transport. Following the approach introduced in \cite{macaulay2020two}, we seek to decompose the force fluctuations into two components. The first component arises from balanced forces on a grain and do not lead to the grain's acceleration. The second component arises from the remaining unbalanced forces, which drives grain acceleration. Figure \ref{fig5}c illustrates this decomposition.

To establish the relative importance of these components, we measured the standard deviation of grain acceleration $\hat{a}$. By sampling over the ensemble, localised transient bursts of grain acceleration \cite{wautier2018micro} are averaged out. We call its normalised counterpart $\A= \hat{a}/ (d/t_i^2)$ \textit{acceleration fluctuation} or \textit{micro-acceleration}, where $t_i = \I/\dot\gamma$ is the inertial time. Figure~\ref{fig5}a shows that acceleration fluctuation is proportional to the inertial number, with only a marginal effect of adhesion:

\be
\A(I,C) \approx 10.5 \I.
\ee

\noindent Accordingly, the acceleration standard deviation scales like: $\hat a \propto \I d/t_i^2 = d\dot \gamma/t_i$, which involves a length scale $d$ and a time scale $\sqrt{t_i/\dot\gamma}$ that are both adhesion independent, and similar to those involved in the diffusion of mass \cite{kharel2017vortices,kharel2018shear}. 

Figure \ref{fig5}b shows the contribution of balanced and unbalanced forces to the force fluctuation. It provides evidence for an approximately linear relation between the force and acceleration fluctuations:

\be \label{eqn:df_A}
\F = \F_0(C) + \kappa(C)\A.
\ee

\begin{figure*}[ht!]
	\centerline{\includegraphics[width=\linewidth]{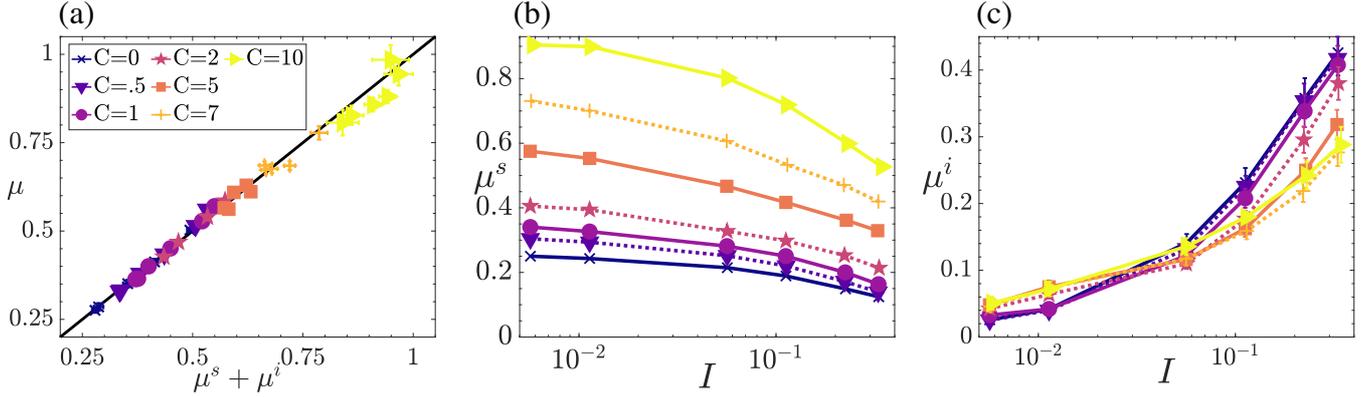}}
	\caption{{\bf Two mechanisms of momentum transport} (constant grain stiffness $K=10^{-3}$). (a) Assessment of the proposed decomposition of the friction $\mu$ into static and inertial components in Eq.~(\ref{eq:muDecomp},\ref{eq:musmui}): markers show the coefficient of friction $\mu$ measured in the flows versus the sum of the components $\mu^s$ and $\mu^i$ calculated from micro-structural quantities as per Eq.~(\ref{eq:musmui}). The line shows the function $\mu= \mu^s+\mu^i$. (b,c) Static and inertial components of the friction calculated by Eq.~(\ref{eq:musmui}). }
	\label{fig6}
\end{figure*}

\noindent We translate this force decomposition into a friction decomposition by combining Eq.~(\ref{eqn:df_A}) and Eq.~(\ref{eqn:mu_df}):

\bee \label{eq:muDecomp}
\mu &=& \mu^s + \mu^i;\\
\mu^s &=& \beta \F_0;\; \mu^i=\kappa \beta \A. \label{eq:musmui}
\eee

\noindent We call $\mu^s$ the \textit{static component} of the friction, as it arises from balanced contact forces only. We call $\mu^i$ the \textit{inertial component} of the friction as it arises from unbalanced forces only, and therefore from grain accelerations. Figure~\ref{fig6}a confirms the validity of this decomposition by comparing the measured friction coefficient $\mu$ to the sum $\mu^s+\mu^i$ as per equation Eq.~(\ref{eq:musmui}). 

Figures~\ref{fig6}b and c show the effect of adhesion on the two components $\mu^s$ and $\mu^i$. The inertial component $\mu^i$ is not greatly influenced by the cohesion strength, and increases approximately linearly with the inertial number. This component drives the shear-rate strengthening effect. In contrast, the static component $\mu^s$ is strongly enhanced by adhesive forces. The enhancement is most pronounced at low inertial numbers. However, increasing the inertial number leads to a reduction in $\mu^s$ for a fixed value of cohesion strength $C$. Accordingly, this component drives the shear-rate weakening effect as cohesive flows transition from the cohesive regime to the non-cohesive regime with increasing shear rate.

\section*{Discussion and conclusion}

%This framework was built from the study of the simplest mode of adhesion (involving a constant attractive force $f_0$ and a Hookean spring), and the simplest flow configuration where shear is steady and homogeneous. 

This study introduced a framework that rationalises the effect of adhesive contact forces on the viscosity of granular materials. The framework includes a new scaling law for the cohesive granular viscosity that can readily be used in continuum fluid mechanics modelling, and some evidence of the micro-mechanical processes controlling it.

The phenomenological scaling law for cohesive granular viscosity that we propose is based on the frictional model Eq.~(\ref{eq:mud}) established for non-cohesive grains. The effect of adhesion is to enhance the friction $\mu$ by a factor $g$ given in Eq.~(\ref{eq:g}). The major advance of this finding is to show that the effect of adhesion on friction is controlled by two dimensionless numbers, namely the cohesion-strength $C$ and the cohesion-energy $W$, which compare the adhesive contact strength and energy to the characteristic force and kinetic energy in the flow. This progresses the commonly accepted view that only one adhesion-related dimensionless number could control the viscosity and reconciles the apparent discrepancies of several existing scaling laws.

In particular, our results indicate that the transition between non-cohesive and cohesive regimes is controlled by these two numbers --- adhesion does not significantly increase the viscosity of granular flows if either $\C$ or $\W$ is less than unity. In this non-cohesive regime, we propose that the stresses and shear rate are sufficient to pull apart adhesive grains by an epitomised process of static or kinetic de-bonding. Reciprocally, the viscosity increases linearly with $\C$ or $\W$ provided that both numbers are greater than $1$. We further define two cohesive regimes, which we call the cohesive-strength (for $\C<\W$) and cohesive-energy (for $\W>\C$) regimes, where the viscosity is controlled by the weakest way that two grains can be pulled apart. \comment{We found that the existence of these two rheological regimes enables a reconciliation between a number of previously proposed rheological models.}

The finding that the cohesion-energy number $\W$ is inversely proportional to the inertial number $\I$ captures the development of a shear-rate weakening behaviour, whereby the friction decreases as the inertial number is increased for a fixed value of $\C$. This behaviour constitutes a promising potential explanation for the existing observations that adhesion induces heterogeneous flows~\cite{alexander2006avalanching, iordanoff2005numerical, klausner2000experimental, gu2014rheology, irani2014impact, yamaguchi2018rheology, kamrin2012nonlocal}. A possible analysis would compare the energy dissipated in a cohesive flow that is homogeneously sheared (uniform friction), to that in a flow featuring a localised shear band (reduced friction) separating two zones with no shear (no dissipation)~\cite{divoux2016shear,degiuli2017friction}. 

We expect this viscous scaling law to apply to more realistic models of adhesion, such as capillary-bridge models, the JKR model or DTM model, which include non-linear Hertz repulsion and a contact area dependent adhesive force. \comment{This expectation is supported by the finding in \cite{roy2016micro} that different models of adhesion lead to the same rheology given that $f_0$ and $w_0$ match.}
We anticipate that the scaling of $\W$ with the grain stiffness $K$ may then involve a slightly different power law, as discussed in the ESI. 

We also expect that this scaling law could be extended to incorporate the effect of the contact dissipation, which was kept constant in our study, possibly by incorporating it into a generalised energy number $W$ in a way similar to ref. \cite{mandal2020insights}. Doing so may explain their remark on the usefulness of the equilibrium grain overlap $\delta_{eq}$, since it is inbuilt into the energy scale $W$.

To further understand the origin of this viscosity scaling law, this study presented an analysis of the internal processes of momentum transport. This led us to pinpoint the micro-structural and micro-dynamical processes affected by adhesion that control the variation in viscosity. These include contact density, force fluctuation, and a measure of force anisotropy. We further developed this micro-mechanical analysis to show that the viscosity arises from two mechanisms of momentum transfer using two distinct pathways: either through balanced contact forces or grain micro-accelerations. This analysis evidenced that adhesion only affects (enhances) momentum transport through balanced contacts. In contrast, it does not significantly affect momentum transport by grain-micro acceleration.

Without cohesion, non-local effects develop at low inertial numbers~\cite{kamrin2019non, kharel2017partial, rognon2015long, zhang2017microscopic, bouzid2013nonlocal}. These effects occur in heterogeneous flows, where the friction law Eq.~(\ref{eq:mud}) becomes affected by the nature of the flow within a typical distance of influence. This distance increases and diverges as the inertial number tends to zero. Little is known on non-locality in cohesive granular flows. Nonetheless, our results indicate that adhesion may strongly affect the prevalent mode of momentum transport at low inertial numbers. This suggests that adhesion may also affect non-local behaviours. The framework introduced in this study may be used as a basis to explore such behaviours.

%%%REFERENCES%%%
\bibliographystyle{plain}
\bibliography{coh_rheo}

\end{document}